# Cluster Decay Half-lives of 5d Transition Metal Nuclei using the Coulomb and Proximity Potential Model

K. E. Abd El Mageed[1], L. I. Abou Salem[1], K.A.Gado[2] Asmaa G. Shalaby[1,3]

*[1]Department of Physics, Faculty of Science, Benha university, 13518 Egypt.*
*[2]Higher Institute of Engineering (BHIE), Bilbes, Sharqia, Egypt.*
*[3]World Laboratory for Cosmology and Particle Physics (WLCAPP), Cairo, Egypt.*

## Abstract

We have applied the Coulomb and Proximity potential model (CPPM), to calculate the half lives for various clusters decay of the selected even-even isotopes of the chosen nuclei. These nuclei are (Hf, W, Os, Pt and Hg) in 5d transition metal region in the periodic table with atomic number $72 \leq Z \leq 80$. Furthermore, half-lives are calculated using the universal formula (UNIV) for cluster decay. The calculated half–lives of alpha decay for the chosen isotopes are in good agreement with the experimental data especially with CPPM results. The alpha and cluster decays are more probable from the parents in the heavier mass number (A=168–180) than from the parents in the lighter mass number (A=156–166).

**Keywords:-** Proximity potential, universal function, half-life time of decay process, *Q*-Value

PACS :- 21.10.Dr, 21.10.Ft, 21.10.Hw

## I. Introduction

The exotic decay or cluster radioactivity is the radioactive decay at which the nuclei emitting particle heavier than α-particle. This cold process is intermediate between α-decay and spontaneous fission. Sandulescu et al. [1] was first predicted this phenomenon on the basis of quantum mechanical fragmentation theory (QMFT). Spontaneous decay of nuclei by emission of clusters heavier than α particle is experimentally established. The emitted $^{14}C$, $^{24, 25, 26}Ne$, $^{28\ 30}Mg$, $^{32\ ,34}Si$ clusters from heavy nuclei were observed and the half- lives measured [2] .

The cluster decay half-lives can be determined theoretically by the one dimensional Wentzel-Kramers Brillouin (WKB) approximation [2] in which the nuclear potential has a significant role. There are many models to calculate the nuclear potential such as the double folding model (DFM) [3, 4] and liquid drop model [5]. In addition, the proximity potential model has been used to study the cluster radioactivity [6–8]. The importance of this model is that, it provides information about the radioactivity of different nuclei [9]. The Coulomb and Proximity Potential model (CPPM) [10] have been used to study alpha and cluster radioactivity in various mass regions of the nuclear chart.

**<u>Emails</u>**
1- karima.abdelmagid@fsc.bu.edu.eg
2- asmaa.shalaby@fsc.bu.edu.eg

The CPPM is used to study the cluster radioactivity and half-life times for various proton rich parents with ($Z$ = 56–64) and ($N$ = 56–72), decaying to doubly magic such as $^{100}Sn$ [10]. This model has been used also to study the cold valleys in the radioactive decay of $^{248-254}Cf$ isotopes and calculated alpha decay half-lives [11]. The cold valleys are the minima in the driving potential (V-$Q$) plots against the mass number of the emitted cluster $A_2$.

We studied the partitions of 4d transition metal nuclei using the core- cluster model in a previous work [12]. In the present work, we attempt to calculate the cluster-decay half-lives of parent nuclei using the coulomb and proximity potential model. We selected the 5d transition metal region in the periodic table in the framework of CPPM. The chosen even-even nuclei for this study are (Hf, W, Os, Pt and Hg) with atomic number $72 \le Z \le 80$.

This paper is organized as follows: the details of the Coulomb and proximity potential model and universal formula are discussed in section II. The results and discussion are represented in section III. Finally, the conclusions from this work is represented.

## II. Theoretical Model

### II.a. The Coulomb and Proximity Potential model (CPPM)

The interaction potential barrier in CPPM is the sum of Coulomb, proximity and centrifugal potentials for the touching configuration and for the separated fragments. Shi and Swiatecki [13] explained the overlap region by a simple power law interpolation. The implication of the proximity potential decreases the height of the potential barrier, which makes the model calculations closely agree with the experimental data.

The total interacting potential barrier for a parent nucleus exhibiting exotic decay is given by

$$V = \frac{Z_1 Z_2 e^2}{r} + V_p(z) + \frac{\hbar^2 \ell(\ell+1)}{2\mu\, r} , \qquad \text{for } z > 0 \qquad (1)$$

Where: $Z_1$ and $Z_2$ are the atomic numbers of the daughter and emitted cluster, $z$ is the distance between the nearby surfaces between daughter and cluster. The distance between fragment centers $r$, $\ell$ is the angular momentum, and $\mu$ is the reduced mass. The proximity potential $V_p(z)$ is represented by Blocki et al. [6],

$$V_p(z) = 4\pi\gamma b \left[\frac{C_1 C_2}{C_1 + C_2}\right] \phi\left(\frac{z}{b}\right) \qquad (2)$$

with the width of the nuclear surface $b \approx 1$, and $\gamma$ is the nuclear surface tension coefficient [14] which is given by

$$\gamma = 0.9517 \left\{ \left(1 - 1.7826\, (N - Z)^2 \right) / A^2 \right\} \qquad \text{MeV/fm}^2 \qquad (3)$$

Where, N, Z and A are the neutron, proton and mass number of the parent respectively. The central radii $C_i$ (i refers to the daughter and/or cluster) related to sharp radii $R_i$ as:

$$C_i = R_i - \left( \frac{b^2}{R_i} \right) \tag{4}$$

To calculate $R_i$ , we can use the semi-empirical formula in terms of mass number $A_i$ [6] as following,

$$R_i = 1.28 A_i^{1/3} + 0.8 A_i^{-1/3} - 0.76 \tag{5}$$

The universal proximity potential [15] φ is given as

$$\phi(\varepsilon) = -4.41 e^{-\frac{\varepsilon}{0.7176}} \qquad \text{for } \varepsilon \geq 1.9475 \tag{6}$$

$$\phi(\varepsilon) = -1.7817 + 0.927\varepsilon + 0.0169\varepsilon^2 - 0.05148\varepsilon^3 \qquad \text{for } 0 \leq \varepsilon \leq 1.9475 \tag{7}$$

With $\varepsilon = z/b$. Using one – dimensional WKB approximation, the barrier penetrability P is given as

$$P = \exp\left( -\frac{2}{\hbar} \int_a^b \sqrt{2\mu(V-Q)}\,dz \right) \tag{8}$$

In eq.(8) the reduced mass $\mu = mA_1A_2/A$, m is the nucleon mass and $A_1$, $A_2$ are the mass numbers of daughter and emitted cluster respectively. The turning points “a” and “b” are determined from V(a) = V(b) = *Q*. The *Q*-value is given by

$$Q = M(A,Z) - M(A_1,Z_1) - M(A_2,Z_2) \tag{9}$$

Where: M (A, Z), M ($A_1$, $Z_1$) and M ($A_2$, $Z_2$) are the atomic masses of the parent, daughter and emitted cluster respectively. The spontaneous cluster decay process occurs when the *Q*-value is positive. The *Q*-values for all the cluster-decay are determined using the experimental mass table [16]. The half-life time of the cluster decay is given by,

$$T_{1/2} = \left( \frac{\ell n 2}{\lambda} \right) = \left( \frac{\ell n 2}{\nu P} \right) \tag{10}$$

Where, $\lambda$ is the decay constant, $\nu$ is the assault frequency written as $\nu = \frac{\omega}{2\pi} = \frac{2E_\nu}{h}$,

and $E_\nu$ is the empirical zero point vibration energy which is given by [17],

$$E_{\nu} = Q\left(0.056 + 0.039 \exp\left[\frac{(4-A_2)}{2.5}\right]\right) \quad , A_2 \geq 4 \tag{11}$$

**II.b The universal Formula (UINV)**

The half–life for the cluster decay is also evaluated by Poenaru et al, [18], in terms of a universal formula.

$$\log_{10} T_{1/2}(S) = -\log_{10} P - \log_{10} S + [\log_{10}(\ell n2) - \log_{10} \nu] \tag{12}$$

Where: P is the penetrability of an external Coulomb barrier and S is the preformation probability of the cluster at the nuclear surface which depends only on the mass number of the emitted cluster $A_2$, and $\nu$ is a constant. The logarithm of the penetrability of an external Coulomb barrier P is given as

$$-\log_{10} P = 0.22873(\mu_A Z_1 Z_2 R_b)^{\frac{1}{2}} \times \left(\arccos\sqrt{r} - \sqrt{r(1-r)}\right) \tag{13}$$

Where: $\mu_A = A_1 A_2 / A$, $r = R_a / R_b$ where $R_a$ and $R_b$ are the turning points. The first turning point $R_a$ is the separation distance at the touching configuration, and can be expressed as, $R_a = R_t = R_1 + R_2$ , Where $R_t = 1.2249\left(A_1^{1/3} + A_2^{1/3}\right)$, and the second turning point defined as $R_b = \frac{e^2 Z_1 Z_2}{Q}$.

The logarithm of the preformation factor S is given as [18]

$$\log_{10} S = -0.598(A_2 - 1) \tag{14}$$

The last two terms in the parentheses in equation (12) is denoted the additive constant $C_{ee}$. In the case of even-even nucleus, $C_{ee}$ is given by,

$$C_{ee} = [-\log_{10} \nu + \log_{10}(\ell n2)] = -22.16917 \tag{15}$$

**III. Results and Discussion**

In the present work we have applied the coulomb and proximity potential model (CPPM) to the cluster-decay of the 5d transition metal region in the periodic table. The selected nuclei (Hf, W, Os, Pt and Hg) have the atomic number in the range $72 \leq Z \leq 80$. This study beginning with the selection of the probable cluster from the selected isotopes through the cold valley plotting, in which (V – *Q*) plotted versus $A_2$. The driving potential (V – *Q*) is calculated for a specific parent for all possible cluster –daughter combinations. The driving potential is defined as the difference between the interaction potential V and the Q-value of the reaction.

Figure (1a-1e) presents the plots of the driving potential versus the mass number of the cluster $A_2$ for the chosen nuclei; this relation is called cold valley plots. From these figures one can notice that, the minima of the driving potential which represent the most probable decay are due to the shell closure of one or both the cluster and daughter. So that, the most probable clusters for the decay process from all the selected nuclei are $^{4}He$, $^{8}Be$, $^{12}C$, $^{16}O$, $^{20,22}Ne$, $^{24,26}Mg$, $^{32}S$, $^{30,32}Si$ and $^{38}Ar$. In addition to that, other deeper minima (valleys) can be found for the same parent nuclei, if one increases the mass number of the cluster.

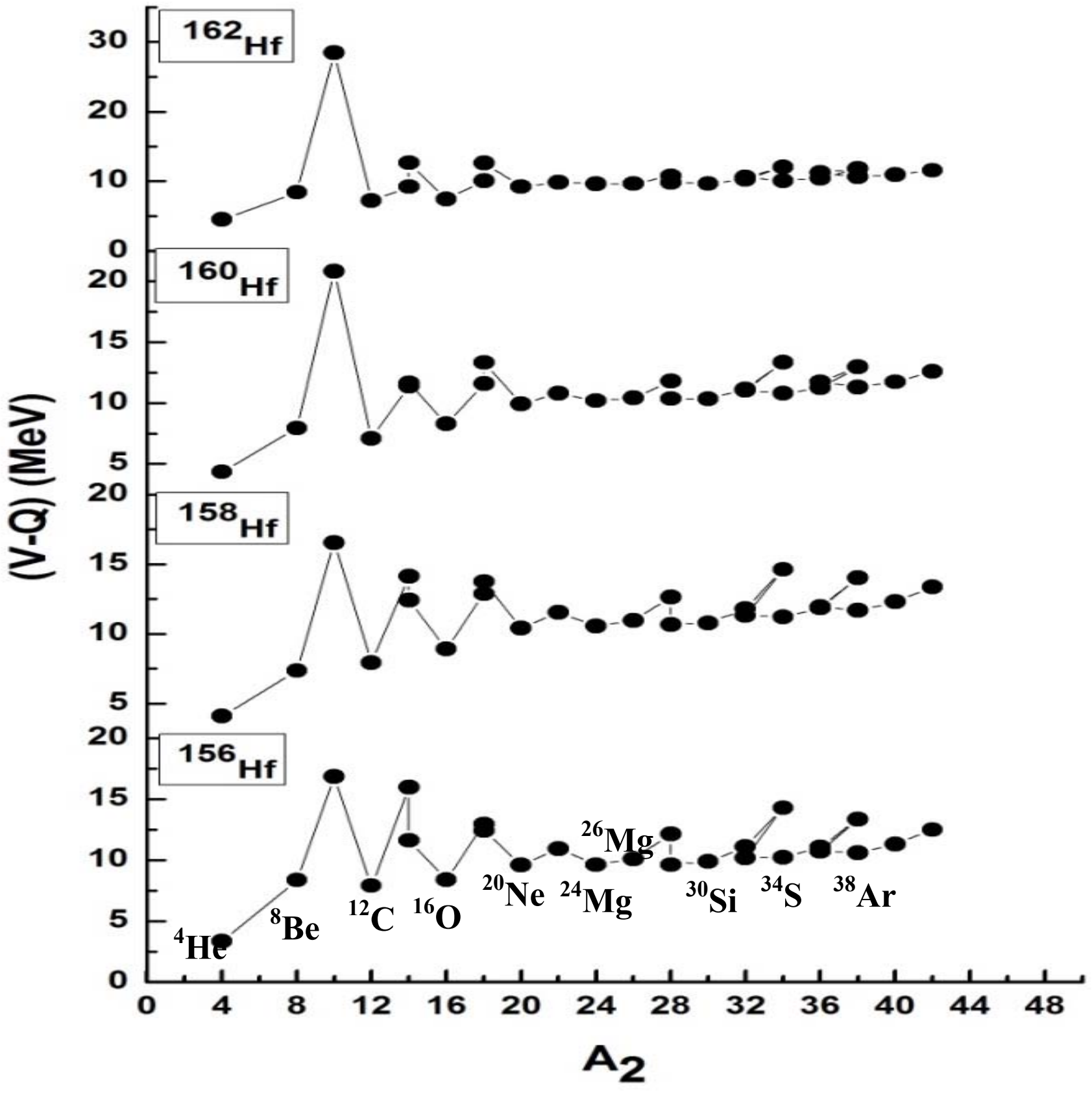

**Figure (1-a) The driving potential (V-$Q$) as a function of the mass number of the emitted cluster $A_2$, for $^{158-162}$Hf isotopes.**

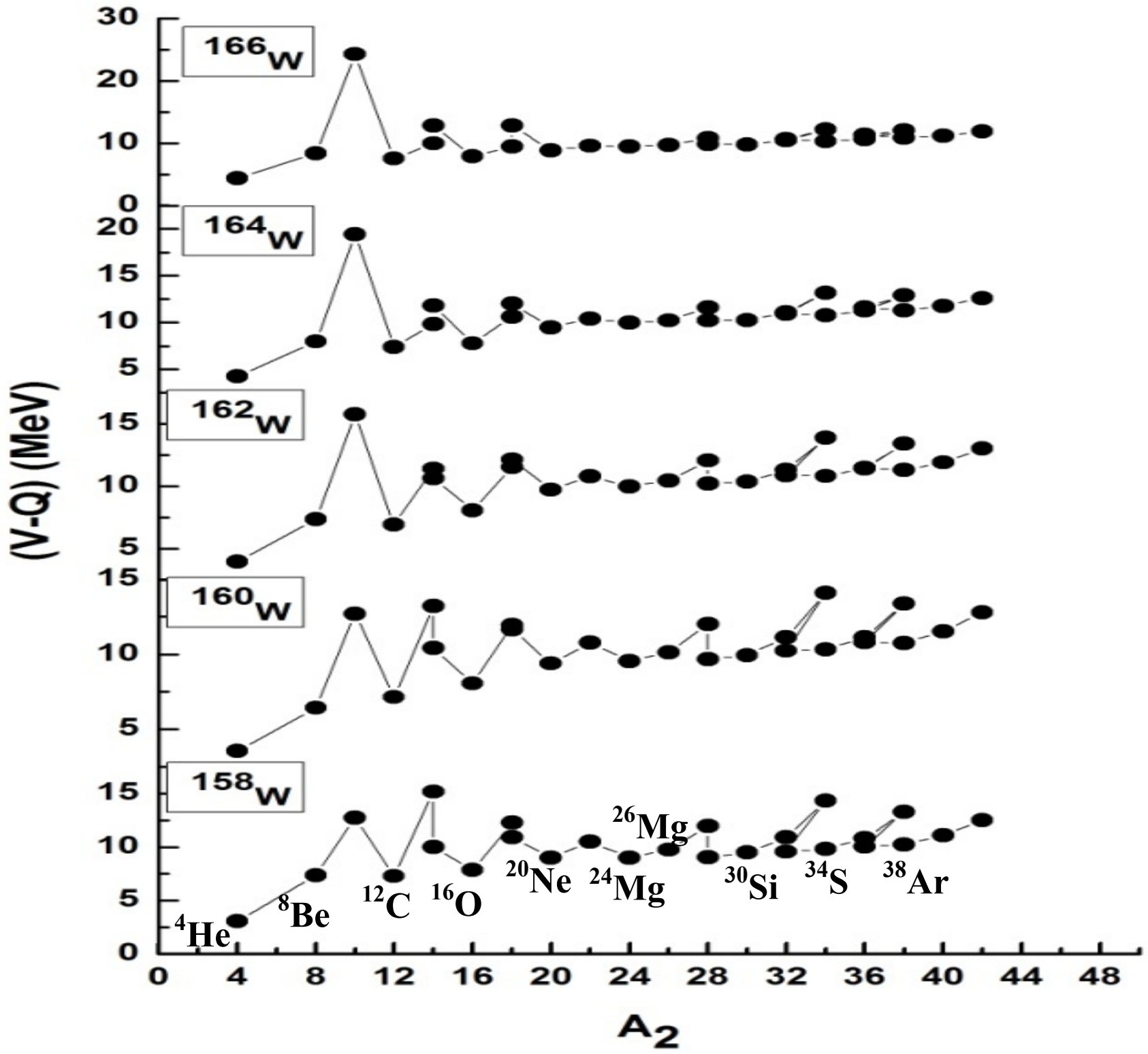

**Figure (1-b) The driving potential (V-*Q*) as a function of the mass number of the emitted cluster $A_2$, for $^{158\text{-}166}$W isotopes.**

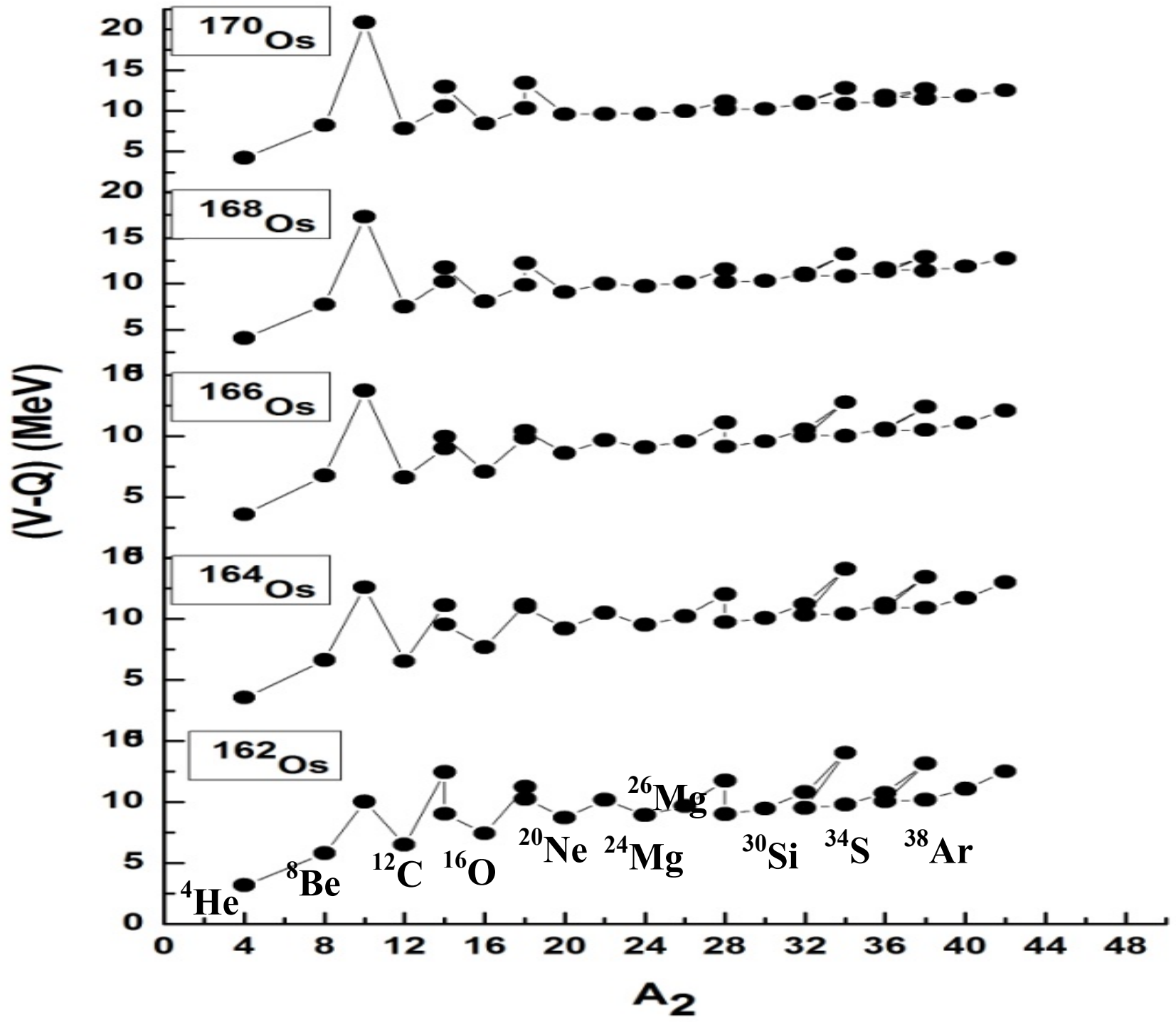

**Figure (1-c) The driving potential (V-*Q*) as a function of the mass number of the emitted cluster $A_2$, for $^{162\text{-}170}$Os isotopes.**

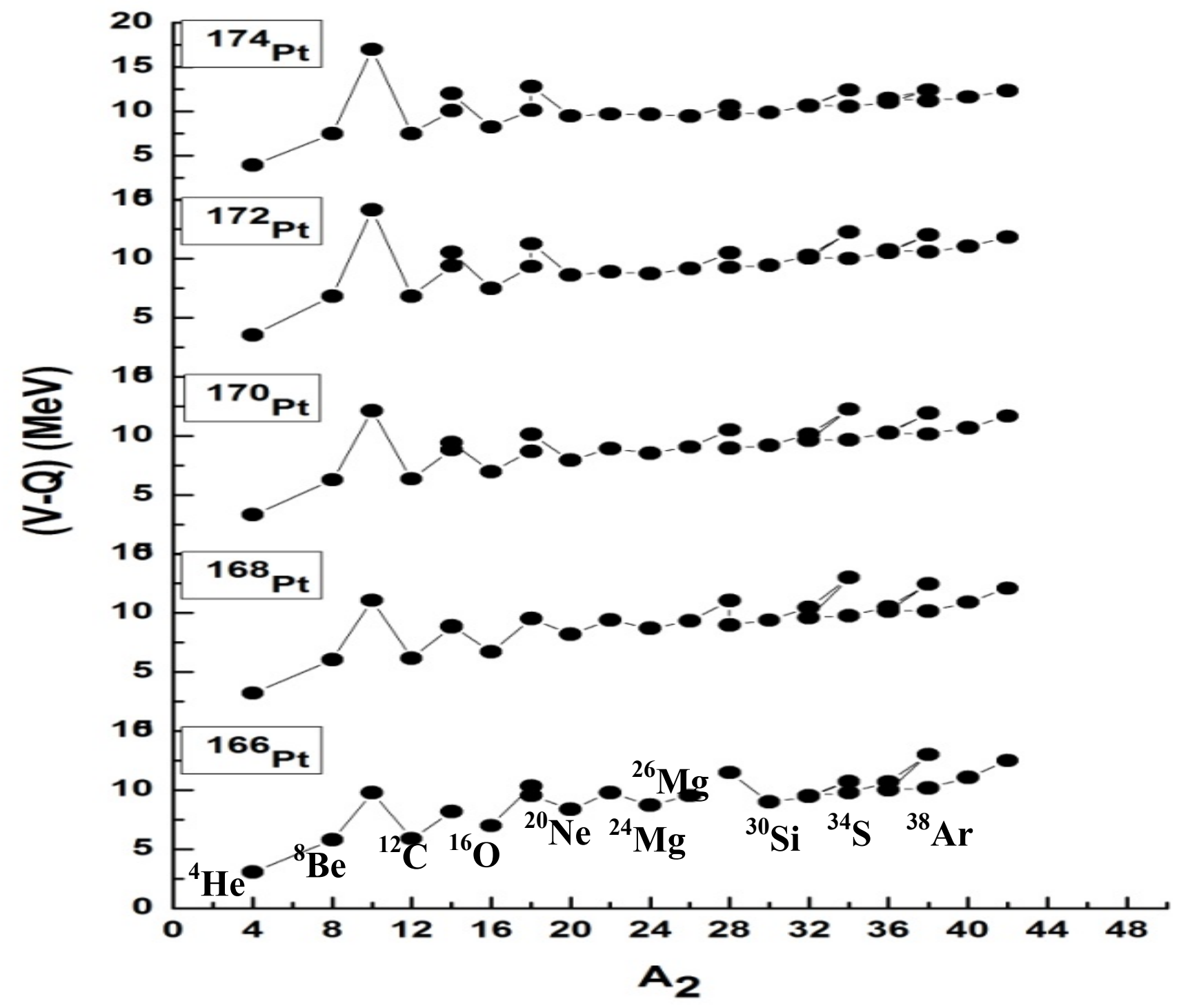

**Figure (1-d) The driving potential (V-*Q*) as a function of the mass number of the emitted cluster $A_2$, for $^{166\text{-}174}$Pt isotopes.**

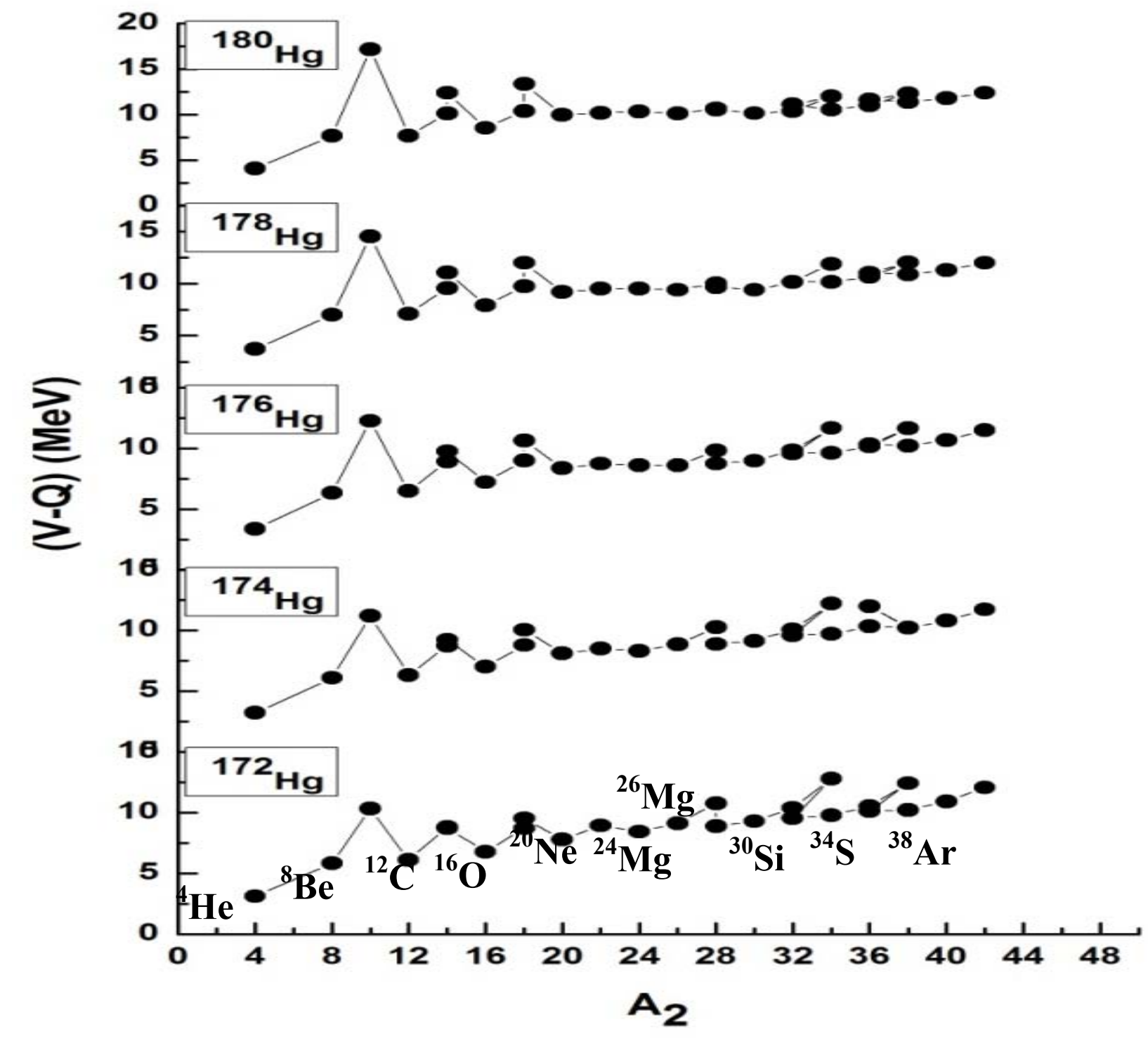

**Figure (1-e) The driving potential (V-*Q*) as a function of the mass number of the emitted $A_2$, for $^{172\text{-}180}$Hg isotopes.**

The values of logarithm of the half-life time, $\log_{10}$ ($T_{1/2}$), for the chosen clusters are calculated using the Coulomb proximity potential (CPPM), and the universal formula (UNIV) from eq.(10) and eq.(12) respectively. The selected isotopes of even-even nuclei are $^{156-162}$Hf, $^{158-166}$W, $^{162-170}$Os, $^{166-174}$Pt, and $^{172-180}$Hg. The calculated values of $\log_{10}$ ($T_{1/2}$) and the available experimental ones [16] for the clusters are listed in table (1). It is obvious that, the calculated values using CPPM are in agreement with the experimental values of $\alpha$-decay more than the UNIV formula. We conclude that, this deviation is due to the difference in potential forms CPPM, and UNIV. Also it is clear that, the Q-value of the clusters have N = Z, is greater than the *Q*-value of the clusters have $N \neq Z$. On the other hand, the values of $\log_{10}$ ($T_{1/2}$), for N = Z are smaller than that of $N \neq Z$. This can be observed clearly, for example in the clusters ($^{20}$Ne, $^{22}$Ne, $^{24}$Mg, $^{26}$Mg) emerged from the decay of $^{156}$Hf and $^{158}$Hf. In comparison for other isotopes this property can be extracted.

The calculated values of $\log_{10}$ ($T_{1/2}$) using CPPM plotted as a function of the mass number of the parents for all the clusters presented in figures (2a-2e). From figure (2a) the values of $\log_{10}$ ($T_{1/2}$) is plotted against the mass number of the isotopes of Hf with different clusters. As the atomic number is fixed Z = 72 for $^{156-162}Hf_{72}$, the plot ($\log_{10}$ ($T_{1/2}$) vs. A) is equivalent to ($\log_{10}$ ($T_{1/2}$) vs. N), in which N is the number of neutrons. From this figure, it is clear that, the lower line is for $^{8}$Be cluster while the upper one for $^{34}$S. Therefore, the $\log_{10}$ ($T_{1/2}$) increases with increasing the mass number of the clusters. This means that, the life time of the decay process from the parent nuclei is longer as the mass number of the clusters is larger. Figures (2b-2e) represent the same plotting mentioned above in the other elements (W, Os, Pt, and Hg).

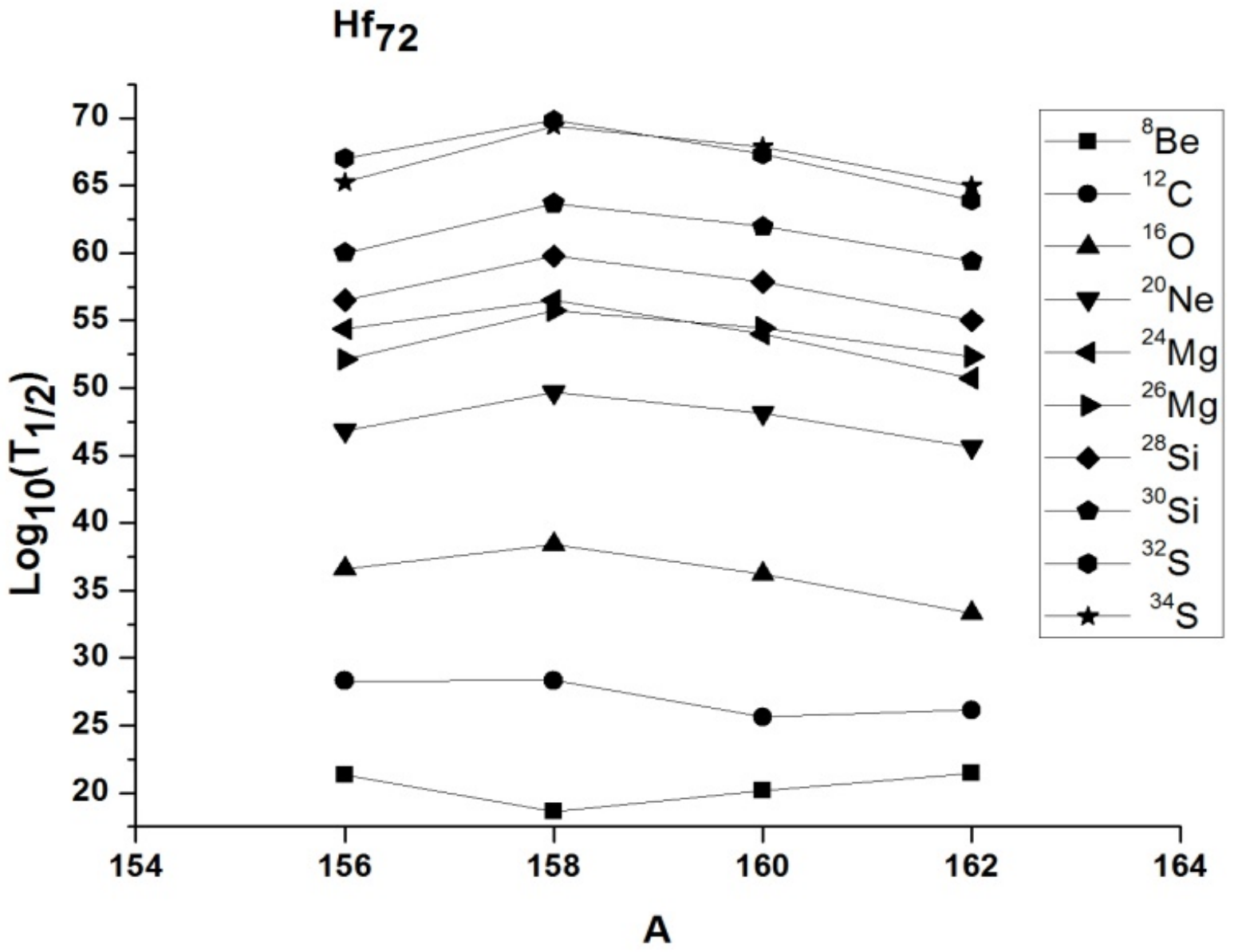

**Figure (2-a) :- The values of $\log_{10}T_{1/2}$ versus the mass number (A) of parent isotopes $^{156-162}$Hf for different clusters.**

**Table (1). :- The values of $Log_{10}(T_{1/2})$ for different isotopes calculated by CPPM and UNIV, in comparison with the available experimental data.**

| Parent Nuclei | Cluster Nuclei | Daughter Nuclei | *Q*-Value (MeV) [16] | $Log_{10}(T_{1/2})$ | | |
|---|---|---|---|---|---|---|
| | | | | The Present Work | | Exp.[ 16 ] |
| | | | | CPPM | UNIV | |
| | $^{4}He$ | $^{152}Yb$ | 6.0285 | -1.62 | -1.61 | -1.64 |
| $^{156}Hf$ | $^{8}Be$ | $^{148}Er$ | 8.6700 | 21.31 | 31.70 | - |
| | $^{12}C$ | $^{144}Dy$ | 18.7030 | 28.32 | 28.62 | - |
| | $^{16}O$ | $^{140}Gd$ | 28.6520 | 36.62 | 30.32 | - |
| | $^{20}Ne$ | $^{136}Sm$ | 35.9860 | 46.83 | 37.84 | - |
| | $^{22}Ne$ | $^{134}Sm$ | 31.5330 | 54.37 | 52.36 | - |
| | $^{24}Mg$ | $^{132}Nd$ | 47.4920 | 52.12 | 35.56 | - |
| | $^{26}Mg$ | $^{130}Nd$ | 44.9440 | 56.52 | 42.78 | - |
| | $^{28}Si$ | $^{128}Ce$ | 59.1590 | 56.79 | 33.54 | - |
| | $^{30}Si$ | $^{126}Ce$ | 57.3860 | 59.99 | 38.18 | - |
| | $^{32}Si$ | $^{124}Ce$ | 51.1260 | 67.03 | 51.70 | - |
| | $^{32}S$ | $^{124}Ba$ | 67.238 | 63.16 | 36.43 | - |
| | $^{34}S$ | $^{122}Ba$ | 66.6730 | 65.25 | 38.93 | - |
| $^{158}Hf$ | $^{4}He$ | $^{154}Yb$ | 5.4047 | 0.45 | 0.92 | 0.45 |
| | $^{8}Be$ | $^{150}Er$ | 10.7871 | 18.61 | 20.82 | - |
| | $^{12}C$ | $^{146}Dy$ | 20.4524 | 28.33 | 23.26 | - |
| | $^{16}O$ | $^{142}Gd$ | 29.5940 | 38.39 | 27.96 | - |
| | $^{20}Ne$ | $^{138}Sm$ | 36.4372 | 49.67 | 36.69 | - |
| | $^{22}Ne$ | $^{136}Sm$ | 32.7331 | 56.51 | 48.78 | - |
| | $^{24}Mg$ | $^{134}Nd$ | 47.4775 | 55.74 | 35.43 | - |
| | $^{26}Mg$ | $^{132}Nd$ | 45.5378 | 59.78 | 41.41 | - |
| | $^{28}Si$ | $^{130}Ce$ | 58.8132 | 60.93 | 33.9 | - |
| | $^{30}Si$ | $^{128}Ce$ | 57.8644 | 63.66 | 37.23 | - |
| | $^{32}Si$ | $^{126}Ce$ | 52.7957 | 69.85 | 48.18 | - |
| | $^{32}S$ | $^{126}Ba$ | 66.5829 | 67.85 | 37.19 | - |
| | $^{34}S$ | $^{124}Ba$ | 66.9190 | 69.41 | 38.39 | - |
| | $^{4}He$ | $^{156}Yb$ | 4.9023 | 1.12 | 3.34 | 1.13 |
| $^{160}Hf$ | $^{8}Be$ | $^{152}Er$ | 9.6210 | 20.17 | 26.27 | - |
| | $^{12}C$ | $^{148}Dy$ | 21.9219 | 25.63 | 19.33 | - |
| | $^{16}O$ | $^{144}Gd$ | 30.5652 | 36.24 | 25.71 | - |
| | $^{20}Ne$ | $^{140}Sm$ | 36.5666 | 48.16 | 36.31 | - |
| | $^{22}Ne$ | $^{138}Sm$ | 33.5912 | 53.99 | 46.30 | - |
| | $^{24}Mg$ | $^{136}Nd$ | 47.2016 | 54.47 | 35.84 | - |
| | $^{26}Mg$ | $^{134}Nd$ | 45.9297 | 57.86 | 40.53 | - |
| | $^{28}Si$ | $^{132}Ce$ | 58.0325 | 59.92 | 34.94 | - |
| | $^{30}Si$ | $^{130}Ce$ | 57.9246 | 61.98 | 37.02 | - |
| | $^{32}Si$ | $^{128}Ce$ | 53.6803 | 67.33 | 38.66 | - |
| | $^{32}S$ | $^{128}Ba$ | 65.4632 | 66.94 | 38.64 | - |
| | $^{34}S$ | $^{126}Ba$ | 66.6703 | 67.85 | 38.66 | - |

**Continue**

| Parent Nuclei | Cluster Nuclei | Daughter Nuclei | *Q*-Value (MeV) [16] | $Log_{10}(T_{1/2})$ The Present Work CPPM | UNIV | Exp.[16] |
|---|---|---|---|---|---|---|
| | $^{4}He$ | $^{158}Yb$ | 4.4162 | 1.58 | 6.05 | 1.59 |
| | $^{8}Be$ | $^{154}Er$ | 8.4942 | 21.45 | 32.57 | - |
| | $^{12}C$ | $^{150}Dy$ | 20.1404 | 26.14 | 23.94 | - |
| | $^{16}O$ | $^{146}Gd$ | 31.6535 | 33.34 | 23.25 | - |
| | $^{20}Ne$ | $^{142}Sm$ | 36.8595 | 45.61 | 35.48 | - |
| $^{162}$**Hf** | $^{22}Ne$ | $^{140}Sm$ | 34.3116 | 50.74 | 44.28 | - |
| | $^{24}Mg$ | $^{138}Nd$ | 46.7828 | 52.34 | 36.40 | - |
| | $^{26}Mg$ | $^{136}Nd$ | 46.2448 | 55.00 | 39.70 | - |
| | $^{28}Si$ | $^{134}Ce$ | 57.1566 | 57.9 | 36.16 | - |
| | $^{30}Si$ | $^{132}Ce$ | 57.7349 | 59.40 | 37.10 | - |
| | $^{32}Si$ | $^{130}Ce$ | 54.3315 | 63.92 | 44.93 | - |
| | $^{32}S$ | $^{130}Ba$ | 64.1082 | 65.03 | 40.55 | - |
| | $^{34}S$ | $^{132}Ba$ | 66.6703 | 64.95 | 38.39 | - |
| $^{158}$**W** | $^{4}He$ | $^{154}Hf$ | 6.6126 | -2.42 | -2.84 | -2.90 |
| | $^{8}Be$ | $^{150}Yb$ | 10.001 | 18.61 | 26.46 | - |
| | $^{12}C$ | $^{146}Er$ | 20.627 | 26.27 | 25.09 | - |
| | $^{16}O$ | $^{142}Dy$ | 31.162 | 34.66 | 27.17 | - |
| | $^{20}Ne$ | $^{138}Gd$ | 39.004 | 44.72 | 34.49 | - |
| | $^{22}Ne$ | $^{136}Gd$ | 33.420 | 52.92 | 50.73 | - |
| | $^{24}Mg$ | $^{134}Sm$ | 51.615 | 49.80 | 31.80 | - |
| | $^{26}Mg$ | $^{132}Sm$ | 47.599 | 55.13 | 41.15 | - |
| | $^{28}Si$ | $^{130}Nd$ | 64.394 | 54.33 | 29.53 | - |
| | $^{30}Si$ | $^{128}Nd$ | 61.052 | 58.42 | 36.10 | - |
| | $^{32}Si$ | $^{126}Nd$ | 53.376 | 66.24 | 51.50 | - |
| | $^{32}S$ | $^{126}Ce$ | 73.141 | 60.63 | 32.29 | - |
| | $^{34}S$ | $^{124}Ce$ | 71.153 | 63.54 | 36.53 | - |
| $^{160}$**W** | $^{4}He$ | $^{156}Hf$ | 6.0655 | -1.06 | 3.55 | -1.00 |
| | $^{8}Be$ | $^{152}Yb$ | 12.002 | 16.02 | 6.44 | - |
| | $^{12}C$ | $^{148}Er$ | 22.102 | 25.82 | 7.15 | - |
| | $^{16}O$ | $^{144}Dy$ | 31.93 | 35.46 | 8.05 | - |
| | $^{20}Ne$ | $^{140}Gd$ | 39.447 | 46.14 | 9.39 | - |
| | $^{22}Ne$ | $^{138}Gd$ | 34.305 | 53.94 | 10.77 | - |
| | $^{24}Mg$ | $^{136}Sm$ | 51.368 | 51.98 | 9.54 | - |
| | $^{26}Mg$ | $^{134}Sm$ | 48.214 | 56.74 | 10.13 | - |
| | $^{28}Si$ | $^{132}Nd$ | 63.542 | 57.05 | 30.52 | - |
| | $^{30}Si$ | $^{130}Nd$ | 61.652 | 60.27 | 9.92 | - |
| | $^{32}Si$ | $^{128}Nd$ | 55.015 | 67.30 | 11.12 | - |
| | $^{32}S$ | $^{128}Ce$ | 72.173 | 63.6 | 33.35 | - |
| | $^{34}S$ | $^{126}Ce$ | 71.375 | 65.85 | 10.32 | - |

Continue

| Parent Nuclei | Cluster Nuclei | Daughter Nuclei | *Q*-Value (MeV) [16] | Log$_{10}$(T$_{1/2}$) | | |
|---|---|---|---|---|---|---|
| | | | | The Present Work | | Exp.[16] |
| | | | | CPPM | UNIV | |
| **$^{162}$W** | $^{4}$He | $^{158}$Hf | 5.6773 | 0.13 | 0.63 | 0.13 |
| | $^{8}$Be | $^{154}$Yb | 10.9901 | 18.68 | 21.72 | - |
| | $^{12}$C | $^{150}$Er | 23.8309 | 25.13 | 16.80 | - |
| | $^{16}$O | $^{146}$Dy | 33.2916 | 35.62 | 22.53 | - |
| | $^{20}$Ne | $^{142}$Gd | 40.0011 | 47.34 | 32.19 | - |
| | $^{22}$Ne | $^{140}$Gd | 35.8067 | 54.15 | 44.17 | - |
| | $^{24}$Mg | $^{138}$Sm | 51.431 | 53.82 | 31.78 | - |
| | $^{26}$Mg | $^{136}$Sm | 49.0251 | 58.05 | 38.13 | - |
| | $^{28}$Si | $^{134}$Nd | 63.1389 | 59.37 | 30.91 | - |
| | $^{30}$Si | $^{132}$Nd | 61.8584 | 62.26 | 34.54 | - |
| | $^{32}$Si | $^{130}$Nd | 56.6736 | 68.20 | 44.91 | - |
| | $^{32}$S | $^{130}$Ce | 71.4381 | 66.33 | 34.14 | - |
| | $^{34}$S | $^{128}$Ce | 71.4653 | 68.10 | 35.73 | - |
| **$^{164}$W** | $^{4}$He | $^{160}$Hf | 5.2785 | 0.83 | 2.40 | 0.79 |
| | $^{8}$Be | $^{156}$Yb | 10.0889 | 20.26 | 25.81 | - |
| | $^{12}$C | $^{152}$Er | 22.2661 | 26.57 | 20.41 | - |
| | $^{16}$O | $^{148}$Dy | 34.3623 | 34.42 | 20.38 | - |
| | $^{20}$Ne | $^{144}$Gd | 40.5736 | 46.53 | 30.94 | - |
| | $^{22}$Ne | $^{142}$Gd | 36.7564 | 52.80 | 41.72 | - |
| | $^{24}$Mg | $^{140}$Sm | 51.1616 | 53.71 | 32.07 | - |
| | $^{26}$Mg | $^{138}$Sm | 49.4844 | 57.33 | 37.13 | - |
| | $^{28}$Si | $^{136}$Nd | 62.4642 | 59.5 | 31.69 | - |
| | $^{30}$Si | $^{134}$Nd | 61.8515 | 61.93 | 34.38 | - |
| | $^{32}$Si | $^{132}$Nd | 57.2756 | 67.35 | 43.67 | - |
| | $^{32}$S | $^{132}$Ce | 70.2586 | 66.77 | 35.55 | - |
| | $^{34}$S | $^{130}$Ce | 71.1267 | 67.99 | 36.01 | - |
| **$^{166}$W** | $^{4}$He | $^{162}$Hf | 4.8560 | 1.27 | 4.53 | 1.28 |
| | $^{8}$Be | $^{158}$Yb | 9.1804 | 21.32 | 30.56 | - |
| | $^{12}$C | $^{154}$Er | 20.7168 | 27.28 | 24.44 | - |
| | $^{16}$O | $^{150}$Dy | 32.1583 | 35.10 | 24.57 | - |
| | $^{20}$Ne | $^{146}$Gd | 41.2393 | 44.44 | 29.54 | - |
| | $^{22}$Ne | $^{144}$Gd | 37.8961 | 50.00 | 38.94 | - |
| | $^{24}$Mg | $^{142}$Sm | 51.0320 | 52.13 | 32.13 | - |
| | $^{26}$Mg | $^{140}$Sm | 49.7823 | 55.33 | 36.44 | - |
| | $^{28}$Si | $^{138}$Nd | 61.623 | 58.28 | 32.74 | - |
| | $^{30}$Si | $^{136}$Nd | 61.7441 | 60.15 | 34.37 | - |
| | $^{32}$Si | $^{134}$Nd | 57.836 | 64.95 | 42.52 | - |

Continue

| Parent Nuclei | Cluster Nuclei | Daughter Nuclei | *Q*-Value (MeV) [16] | $Log_{10}(T_{1/2})$ The Present Work CPPM | $Log_{10}(T_{1/2})$ The Present Work UNIV | $Log_{10}(T_{1/2})$ Exp.[16] |
|---|---|---|---|---|---|---|
| | $^{32}$S | $^{134}$Ce | 68.9603 | 65.69 | | - |
| | $^{34}$S | $^{132}$Ce | 70.5145 | 66.42 | 37.18 | - |
| | | | | | 36.67 | |
| **$^{162}$Os** | $^{4}$He | $^{158}$W | 6.7673 | -2.20 | -2.59 | -2.67 |
| | $^{8}$Be | $^{154}$Hf | 13.288 | 14.09 | 14.89 | - |
| | $^{12}$C | $^{150}$Yb | 24.135 | 23.73 | 18.16 | - |
| | $^{16}$O | $^{146}$Er | 34.556 | 33.23 | 22.58 | - |
| | $^{20}$Ne | $^{142}$Dy | 42.659 | 43.58 | 30.21 | - |
| | $^{22}$Ne | $^{140}$Dy | 36.352 | 51.90 | 46.42 | - |
| | $^{24}$Mg | $^{138}$Gd | 55.088 | 49.57 | 29.31 | - |
| | $^{26}$Mg | $^{136}$Gd | 50.802 | 54.91 | 38.50 | - |
| | $^{28}$Si | $^{134}$Sm | 68.366 | 54.4 | 27.42 | - |
| | $^{30}$Si | $^{132}$Sm | 65.009 | 58.40 | 33.62 | - |
| | $^{32}$Si | $^{130}$Sm | 57.081 | 66.05 | 48.39 | - |
| | $^{32}$S | $^{130}$Nd | 78.109 | 60.57 | 29.58 | - |
| | $^{34}$S | $^{128}$Nd | 75.743 | 63.65 | 34.03 | - |
| **$^{164}$Os** | $^{4}$He | $^{160}$W | 6.4794 | -1.12 | -1.62 | -1.67 |
| | $^{8}$Be | $^{158}$Hf | 12.453 | 16.53 | 17.73 | - |
| | $^{12}$C | $^{152}$Yb | 25.848 | 23.63 | 14.44 | - |
| | $^{16}$O | $^{148}$Er | 35.743 | 34.13 | 20.32 | - |
| | $^{20}$Ne | $^{144}$Dy | 43.139 | 45.47 | 29.24 | - |
| | $^{24}$Mg | $^{140}$Gd | 55.243 | 51.99 | 28.97 | - |
| | $^{26}$Mg | $^{138}$Gd | 51.399 | 57.14 | 37.34 | - |
| | $^{28}$Si | $^{136}$Sm | 67.831 | 57.47 | 27.93 | - |
| | $^{30}$Si | $^{134}$Sm | 65.336 | 61.05 | 33.04 | - |
| | $^{32}$S | $^{132}$Nd | 76.969 | 64.24 | 30.75 | - |
| | $^{34}$S | $^{130}$Nd | 76.055 | 66.56 | 33.5 | - |
| **$^{166}$Os** | $^{4}$He | $^{162}$W | 6.1386 | -0.94 | -0.39 | -0.70 |
| | $^{8}$Be | $^{160}$Hf | 11.724 | 16.99 | 20.44 | - |
| | $^{12}$C | $^{154}$Yb | 24.4953 | 24.06 | 17.19 | - |
| | $^{16}$O | $^{150}$Er | 37.1315 | 32.01 | 17.8 | - |
| | $^{20}$Ne | $^{146}$Dy | 44.1601 | 43.43 | 27.31 | - |
| | $^{24}$Mg | $^{142}$Gd | 55.4563 | 50.49 | 28.49 | - |
| | $^{26}$Mg | $^{140}$Gd | 52.56 | 54.79 | 35.17 | - |
| | $^{28}$Si | $^{138}$Sm | 67.5538 | 64.59 | 36.4 | - |
| | $^{30}$Si | $^{136}$Sm | 65.8071 | 59.24 | 32.21 | - |
| | $^{32}$S | $^{134}$Nd | 76.2252 | 63.12 | 31.47 | - |
| | $^{34}$S | $^{132}$Nd | 75.9207 | 65.05 | 33.49 | - |

Continue

| Parent Nuclei | Cluster Nuclei | Daughter Nuclei | *Q*-Value (MeV) [16] | $Log_{10}(T_{1/2})$ | | |
|---|---|---|---|---|---|---|
| | | | | The Present Work | | Exp.[16] |
| | | | | CPPM | UNIV | |
| **$^{168}$Os** | $^{4}$He | $^{64}$W | 5.8161 | 0.34 | 0.87 | 0.32 |
| | $^{8}$Be | $^{160}$Hf | 11.0027 | 19.64 | 23.37 | - |
| | $^{12}$C | $^{156}$Yb | 23.2716 | 26.95 | 19.88 | - |
| | $^{16}$O | $^{152}$Er | 35.244 | 35.49 | 20.93 | - |
| | $^{20}$Ne | $^{148}$Dy | 44.9083 | 45.19 | 25.85 | - |
| | $^{22}$Ne | $^{146}$Dy | 40.5927 | 51.46 | 36.30 | - |
| | $^{24}$Mg | $^{144}$Gd | 55.7062 | 52.96 | 27.92 | - |
| | $^{26}$Mg | $^{142}$Gd | 53.1872 | 57.11 | 33.90 | - |
| | $^{28}$Si | $^{140}$Sm | 66.9618 | 59.47 | 28.71 | - |
| | $^{30}$Si | $^{138}$Sm | 65.9438 | 62.16 | 31.79 | - |
| | $^{32}$Si | $^{136}$Sm | 60.9017 | 67.76 | 41.21 | - |
| | $^{32}$S | $^{136}$Nd | 75.2279 | 66.93 | 32.53 | - |
| | $^{34}$S | $^{134}$Nd | 75.5912 | 72.59 | 40.03 | - |
| **$^{170}$Os** | $^{4}$He | $^{166}$W | 5.5368 | 0.86 | 2.06 | 0.86 |
| | $^{8}$Be | $^{162}$Hf | 10.3009 | 20.98 | 26.58 | - |
| | $^{12}$C | $^{158}$Yb | 22.0838 | 28.23 | 22.77 | - |
| | $^{16}$O | $^{154}$Er | 33.4155 | 46.61 | 24.32 | - |
| | $^{20}$Ne | $^{150}$Dy | 42.4249 | 47.18 | 30.08 | - |
| | $^{22}$Ne | $^{148}$Dy | 41.9515 | 50.17 | 33.42 | - |
| | $^{24}$Mg | $^{146}$Gd | 56.0927 | 52.65 | 27.20 | - |
| | $^{26}$Mg | $^{144}$Gd | 54.0476 | 56.41 | 32.34 | - |
| | $^{28}$Si | $^{142}$Sm | 66.553 | 59.71 | 29.07 | - |
| | $^{30}$Si | $^{140}$Sm | 65.9625 | 62.12 | 31.60 | - |
| | $^{32}$Si | $^{138}$Sm | 61.649 | 67.17 | 39.83 | - |
| | $^{32}$S | $^{138}$Nd | 74.1074 | 67.63 | 33.78 | - |
| | $^{34}$S | $^{136}$Nd | 75.2045 | 68.77 | 33.98 | - |

Continue

| Parent Nuclei | Cluster Nuclei | Daughter Nuclei | *Q*-Value (MeV) [16] | $Log_{10}(T_{1/2})$ | | |
|---|---|---|---|---|---|---|
| | | | | The Present Work | | Exp.[16] |
| | | | | CPPM | UNIV | |
| $^{166}$**Pt** | $^{4}$He | $^{162}$Os | 7.2858 | -2.48 | -3.49 | -3.50 |
| | $^{8}$Be | $^{158}$W | 13.961 | 14.07 | 14.25 | - |
| | $^{12}$C | $^{154}$Hf | 27.941 | 21.5 | 12.19 | - |
| | $^{16}$O | $^{150}$Yb | 38.583 | 31.57 | 17.67 | - |
| | $^{20}$Ne | $^{146}$Er | 46.572 | 42.45 | 26.1 | - |
| | $^{24}$Mg | $^{142}$Dy | 59.262 | 48.96 | 26.13 | - |
| | $^{26}$Mg | $^{140}$Dy | 54.253 | 54.59 | 35.74 | - |
| | $^{28}$Si | $^{138}$Gd | 72.358 | 54.5 | 25.46 | - |
| | $^{30}$Si | $^{136}$Gd | 68.731 | 58.58 | 31.72 | - |
| | $^{32}$S | $^{134}$Sm | 82.6 | 60.84 | 27.63 | - |
| | $^{34}$S | $^{132}$Sm | 80.219 | 63.89 | 31.93 | - |
| | $^{36}$Ar | $^{130}$Nd | 92.036 | 66.66 | 29.85 | - |
| $^{168}$**Pt** | $^{4}$He | $^{164}$Os | 6.9896 | -2.1 | -2.59 | -2.69 |
| | $^{8}$Be | $^{160}$W | 13.377 | 14.81 | 16.07 | - |
| | $^{12}$C | $^{156}$Hf | 26.809 | 22.4 | 14.23 | - |
| | $^{16}$O | $^{152}$Yb | 40 | 30.52 | 15.33 | - |
| | $^{20}$Ne | $^{148}$Er | 47.463 | 41.76 | 24.54 | - |
| | $^{24}$Mg | $^{144}$Dy | 59.446 | 48.8 | 25.72 | - |
| | $^{26}$Mg | $^{142}$Dy | 55.276 | 53.8 | 33.9 | - |
| | $^{28}$Si | $^{140}$Gd | 72.217 | 54.56 | 25.45 | - |
| | $^{30}$Si | $^{138}$Gd | 69.032 | 58.37 | 31.15 | - |
| | $^{32}$S | $^{136}$Sm | 81.768 | 61.28 | 28.37 | - |
| | $^{34}$S | $^{134}$Sm | 80.25 | 63.86 | 31.71 | - |
| | $^{36}$Ar | $^{132}$Nd | 90.599 | 67.43 | 31.17 | - |
| $^{170}$**Pt** | $^{4}$He | $^{166}$Os | 6.7073 | -1.71 | -1.68 | -1.86 |
| | $^{8}$Be | $^{162}$W | 12.7541 | 15.67 | 18.16 | - |
| | $^{12}$C | $^{158}$Hf | 25.798 | 23.25 | 16.17 | - |
| | $^{16}$O | $^{154}$Yb | 38.3646 | 31.64 | 17.78 | - |
| | $^{20}$Ne | $^{150}$Er | 48.5687 | 40.95 | 22.7 | - |
| | $^{24}$Mg | $^{146}$Dy | 60.1839 | 48.3 | 24.57 | - |
| | $^{26}$Mg | $^{144}$Dy | 56.4801 | 52.9 | 31.84 | - |
| | $^{28}$Si | $^{142}$Gd | 72.1478 | 54.57 | 25.36 | - |
| | $^{30}$Si | $^{140}$Gd | 69.9107 | 57.8 | 29.83 | - |
| | $^{32}$S | $^{138}$Sm | 81.2088 | 61.58 | 28.81 | - |
| | $^{34}$S | $^{136}$Sm | 80.438 | 63.75 | 31.3 | - |
| | $^{36}$Ar | $^{134}$Nd | 89.5734 | 67.99 | 32.09 | - |

Continue

| Parent Nuclei | Cluster Nuclei | Daughter Nuclei | Q-Value (MeV) [16] | Log$_{10}$(T$_{1/2}$) | | |
|---|---|---|---|---|---|---|
| | | | | The Present Work | | Exp.[16] |
| | | | | CPPM | UNIV | |
| **$^{172}$Pt** | $^{4}$He | $^{168}$Os | 6.4645 | -1.03 | -0.85 | -1.00 |
| | $^{8}$Be | $^{164}$W | 12.1887 | 17.14 | 20.2 | - |
| | $^{12}$C | $^{160}$Hf | 24.8338 | 24.89 | 18.16 | - |
| | $^{16}$O | $^{156}$Yb | 36.898 | 33.65 | 20.14 | - |
| | $^{20}$Ne | $^{152}$Er | 46.4384 | 43.53 | 25.9 | - |
| | $^{24}$Mg | $^{148}$Dy | 60.6893 | 49.18 | 23.75 | - |
| | $^{26}$Mg | $^{146}$Dy | 57.672 | 53.34 | 29.87 | - |
| | $^{28}$Si | $^{144}$Gd | 72.1548 | 55.89 | 25.19 | - |
| | $^{30}$Si | $^{142}$Gd | 70.295 | 58.95 | 29.16 | - |
| | $^{32}$S | $^{140}$Sm | 80.374 | 63.52 | 29.59 | - |
| | $^{34}$S | $^{138}$Sm | 80.332 | 65.32 | 31.24 | - |
| | $^{36}$Ar | $^{136}$Nd | 88.3333 | 70.29 | 33.28 | - |
| **$^{174}$Pt** | $^{4}$He | $^{170}$Os | 6.1831 | -0.09 | 0.19 | -0.05 |
| | $^{8}$Be | $^{166}$W | 11.628 | 19.01 | 22.39 | - |
| | $^{12}$C | $^{162}$Hf | 23.8506 | 27.02 | 20.33 | - |
| | $^{16}$O | $^{158}$Yb | 35.4288 | 36.23 | 22.68 | - |
| | $^{20}$Ne | $^{154}$Er | 44.3284 | 46.81 | 29.35 | - |
| | $^{24}$Mg | $^{150}$Dy | 57.9246 | 52.8 | 27.38 | - |
| | $^{26}$Mg | $^{148}$Dy | 58.7493 | 54.47 | 28.14 | - |
| | $^{28}$Si | $^{146}$Gd | 72.2599 | 57.81 | 24.9 | - |
| | $^{30}$Si | $^{144}$Gd | 70.874 | 60.65 | 28.26 | - |
| | $^{32}$S | $^{142}$Sm | 79.6837 | 66.13 | 30.22 | - |
| | $^{34}$S | $^{140}$Sm | 80.0692 | 67.73 | 31.37 | - |
| | $^{36}$Ar | $^{138}$Nd | 86.9314 | 73.5 | 34.7 | - |

Continue

| Parent Nuclei | Cluster Nuclei | Daughter Nuclei | *Q*-Value (MeV) [16] | Log$_{10}$(T$_{1/2}$) | | |
|---|---|---|---|---|---|---|
| | | | | The Present Work | | Exp.[16] |
| | | | | CPPM | UNIV | |
| **$^{172}$Hg** | $^{4}$He | $^{168}$Pt | 7.5237 | -2.39 | -3.53 | -3.60 |
| | $^{8}$Be | $^{164}$Os | 14.422 | 14.29 | 14.21 | - |
| | $^{12}$C | $^{160}$W | 28.267 | 22.3 | 13.09 | - |
| | $^{16}$O | $^{156}$Hf | 41.495 | 30.9 | 14.98 | - |
| | $^{20}$Ne | $^{152}$Yb | 52.253 | 40.31 | 19.84 | - |
| | $^{24}$Mg | $^{148}$Er | 64.303 | 47.91 | 22.01 | - |
| | $^{28}$Si | $^{144}$Dy | 76.953 | 54.29 | 22.82 | - |
| | $^{30}$Si | $^{142}$Dy | 73.444 | 58.19 | 28.51 | - |
| | $^{32}$S | $^{140}$Gd | 86.688 | 61.34 | 26.06 | - |
| | $^{34}$S | $^{138}$Gd | 84.479 | 64.27 | 30 | - |
| | $^{36}$Ar | $^{136}$Sm | 95.933 | 67.68 | 28.93 | - |
| | $^{38}$Ar | $^{134}$Sm | 94.981 | 69.76 | 31.37 | - |
| | $^{40}$Ar | $^{132}$Sm | 89.009 | 74.6 | 39.59 | - |
| **$^{174}$Hg** | $^{4}$He | $^{170}$Pt | 7.2332 | -2.03 | -2.67 | - 2.67 |
| | $^{8}$Be | $^{166}$Os | 13.8487 | 15 | 15.94 | - |
| | $^{12}$C | $^{162}$W | 27.3539 | 23.02 | 14.73 | - |
| | $^{16}$O | $^{158}$Hf | 40.1931 | 31.76 | 16.82 | - |
| | $^{20}$Ne | $^{154}$Yb | 50.3276 | 41.53 | 22.42 | - |
| | $^{24}$Mg | $^{150}$Er | 65.1184 | 47.4 | 20.87 | - |
| | $^{28}$Si | $^{146}$Dy | 76.953 | 54.26 | 22.65 | - |
| | $^{30}$Si | $^{144}$Dy | 74.3566 | 57.63 | 27.24 | - |
| | $^{32}$S | $^{142}$Gd | 86.3286 | 61.51 | 26.26 | - |
| | $^{34}$S | $^{140}$Gd | 85.0675 | 63.94 | 29.16 | - |
| | $^{36}$Ar | $^{138}$Sm | 95.0829 | 68.12 | 29.59 | - |
| | $^{38}$Ar | $^{136}$Sm | 94.8793 | 69.82 | 31.27 | - |
| | $^{40}$Ar | $^{134}$Sm | 89.769 | 74.18 | 38.47 | - |
| **$^{176}$Hg** | $^{4}$He | $^{172}$Pt | 6.8993 | -1.58 | -1.61 | -1.69 |
| | $^{8}$Be | $^{168}$Os | 13.2719 | 15.76 | 17.81 | - |
| | $^{12}$C | $^{164}$W | 26.4546 | 23.76 | 16.44 | - |
| | $^{16}$O | $^{160}$Hf | 38.895 | 32.65 | 18.78 | - |
| | $^{20}$Ne | $^{156}$Yb | 48.5271 | 42.72 | 25.01 | - |
| | $^{24}$Mg | $^{152}$Er | 62.6542 | 48.82 | 23.76 | - |
| | $^{28}$Si | $^{148}$Dy | 77.5727 | 53.9 | 21.82 | - |
| | $^{30}$Si | $^{146}$Dy | 75.2146 | 57.11 | 26.06 | - |
| | $^{32}$S | $^{144}$Gd | 86.0018 | 61.67 | 26.42 | - |
| | $^{34}$S | $^{142}$Gd | 85.1179 | 63.9 | 28.92 | - |
| | $^{36}$Ar | $^{140}$Sm | 93.9142 | 68.73 | 30.61 | - |
| | $^{38}$Ar | $^{138}$Sm | 94.4393 | 70.06 | 31.54 | - |
| | $^{40}$Ar | $^{136}$Sm | 90.0775 | 74.01 | 37.91 | - |

Continue

| Parent Nuclei | Cluster Nuclei | Daughter Nuclei | *Q*-Value (MeV) [16] | $Log_{10}(T_{1/2})$ | | |
|---|---|---|---|---|---|---|
| | | | | The Present Work | | Exp.[16] |
| | | | | CPPM | UNIV | |
| **$^{178}$Hg** | $^{4}$He | $^{174}$Pt | 6.5773 | -0.57 | -0.5 | -0.57 |
| | $^{8}$Be | $^{170}$Os | 12.6685 | 17.66 | 19.92 | - |
| | $^{12}$C | $^{166}$W | 25.5719 | 25.81 | 18.22 | - |
| | $^{16}$O | $^{162}$Hf | 37.5898 | 35.15 | 20.88 | - |
| | $^{20}$Ne | $^{158}$Yb | 46.7359 | 45.83 | 27.77 | - |
| | $^{24}$Mg | $^{154}$Er | 60.2222 | 52.36 | 26.85 | - |
| | $^{28}$Si | $^{150}$Dy | 74.486 | 57.79 | 25.08 | - |
| | $^{30}$Si | $^{148}$Dy | 75.9699 | 58.92 | 25.02 | - |
| | $^{32}$S | $^{146}$Gd | 85.7848 | 64.18 | 26.47 | - |
| | $^{34}$S | $^{144}$Gd | 85.3749 | 66.22 | 28.45 | - |
| | $^{36}$Ar | $^{142}$Sm | 92.9019 | 71.9 | 31.49 | - |
| | $^{38}$Ar | $^{140}$Sm | 93.8545 | 73.04 | 31.96 | - |
| | $^{40}$Ar | $^{138}$Sm | 90.2214 | 76.71 | 37.54 | - |
| **$^{180}$Hg** | $^{4}$He | $^{176}$Pt | 6.2584 | 0.39 | 0.68 | 0.41 |
| | $^{8}$Be | $^{172}$Os | 12.0516 | 19.49 | 22.26 | - |
| | $^{12}$C | $^{168}$W | 24.6425 | 27.74 | 20.23 | - |
| | $^{16}$O | $^{164}$Hf | 36.3049 | 37.43 | 23.09 | - |
| | $^{20}$Ne | $^{160}$Yb | 44.9565 | 48.71 | 30.72 | - |
| | $^{24}$Mg | $^{156}$Er | 57.8936 | 55.58 | 30.04 | - |
| | $^{28}$Si | $^{152}$Dy | 71.3604 | 61.45 | 28.68 | - |
| | $^{30}$Si | $^{150}$Dy | 73.492 | 62.21 | 27.77 | - |
| | $^{32}$S | $^{148}$Gd | 82.0345 | 68.26 | 30.47 | - |
| | $^{34}$S | $^{146}$Gd | 85.7669 | 67.98 | 27.84 | - |
| | $^{36}$Ar | $^{144}$Sm | 91.9466 | 74.56 | 32.33 | - |
| | $^{38}$Ar | $^{142}$Sm | 93.4511 | 75.43 | 32.2 | - |
| | $^{40}$Ar | $^{140}$Sm | 90.2455 | 78.93 | 37.31 | - |

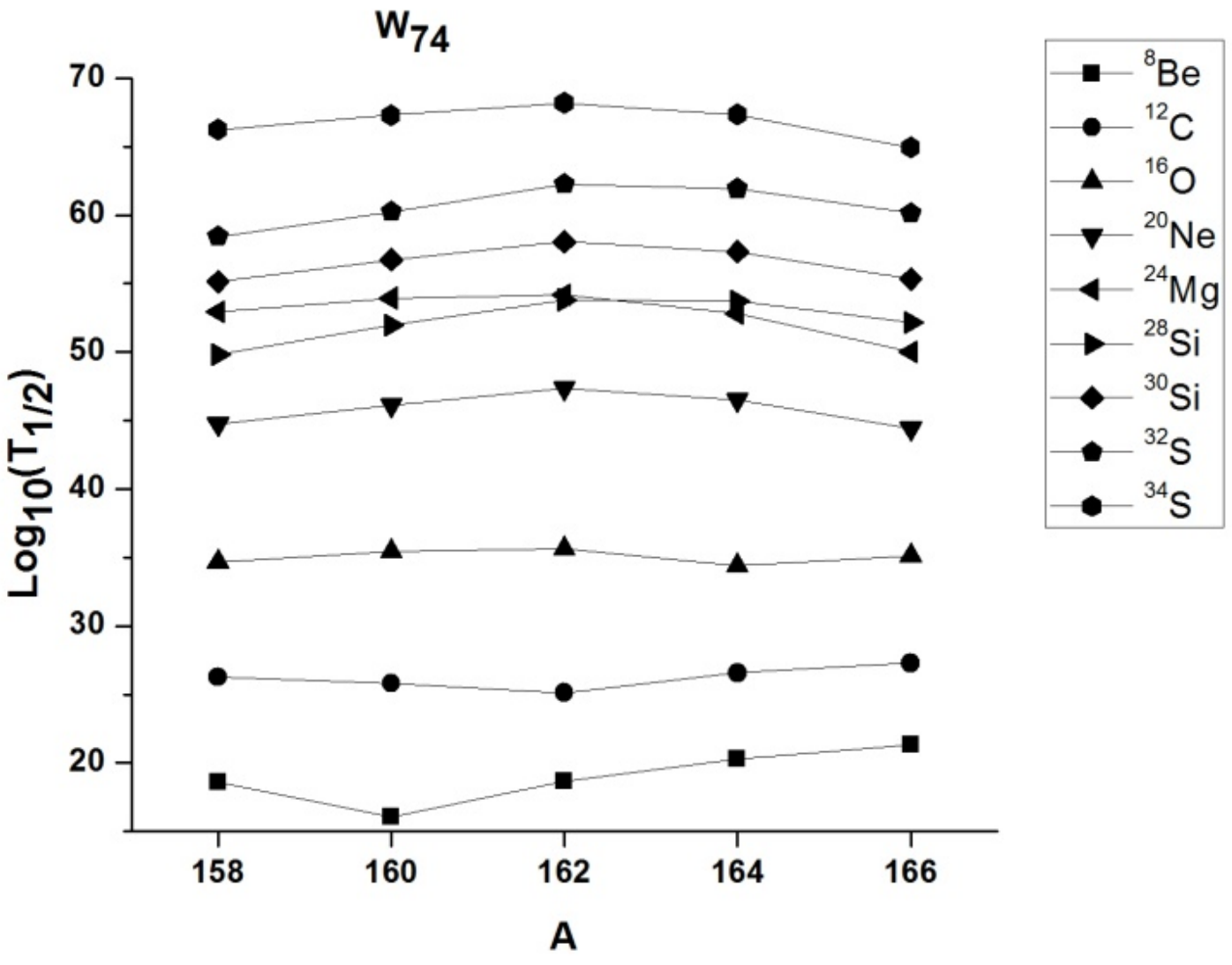

**Figure (2-b) :- The values of $\log_{10}T_{1/2}$ versus the atomic mass number (A) of parent isotopes $^{158\text{-}166}$W for different clusters.**

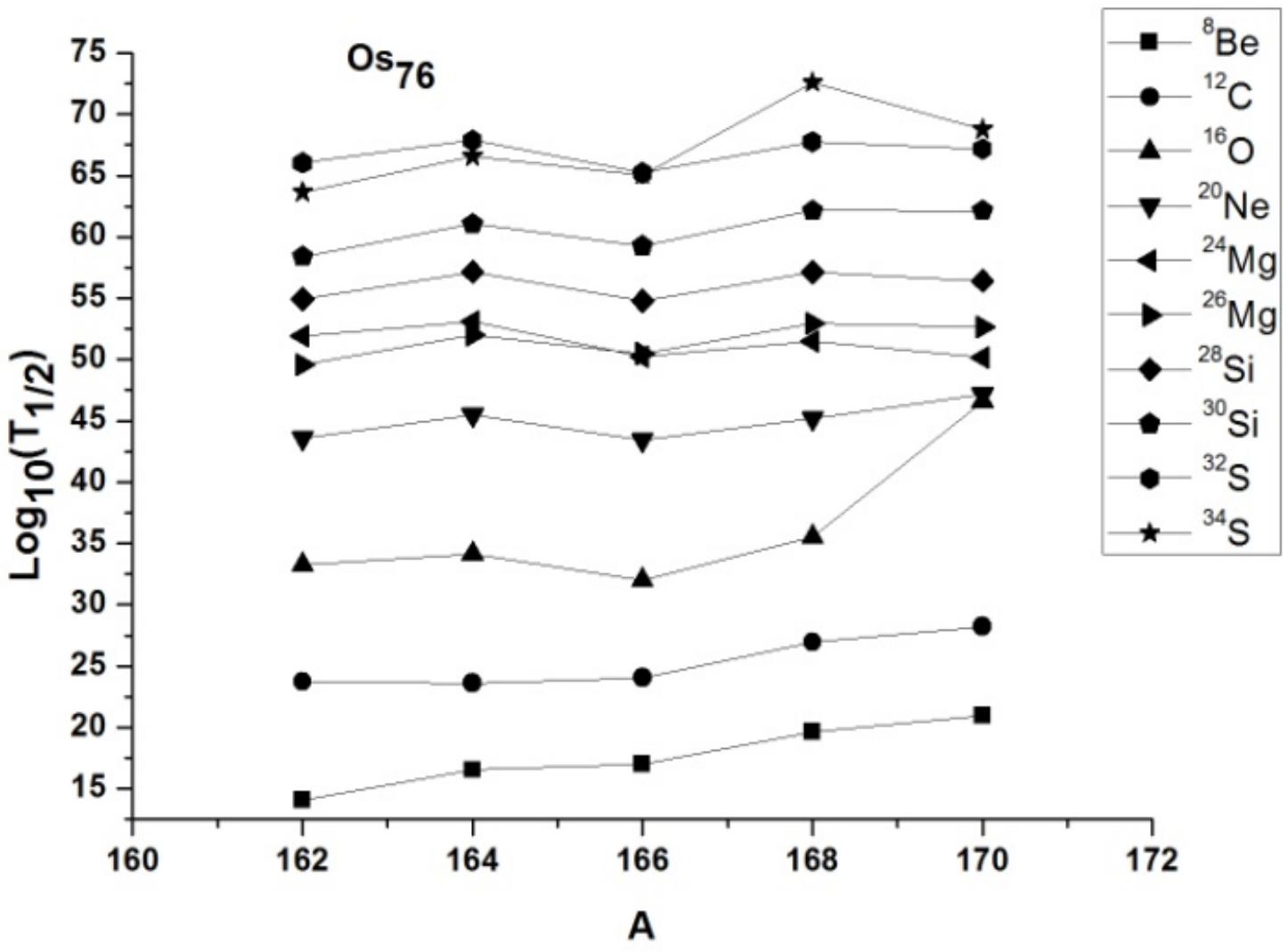

**Figure (2-c) :- The values of $\log_{10}T_{1/2}$ versus the mass number (A) of parent isotopes $^{162\text{-}170}$Os for different clusters.**

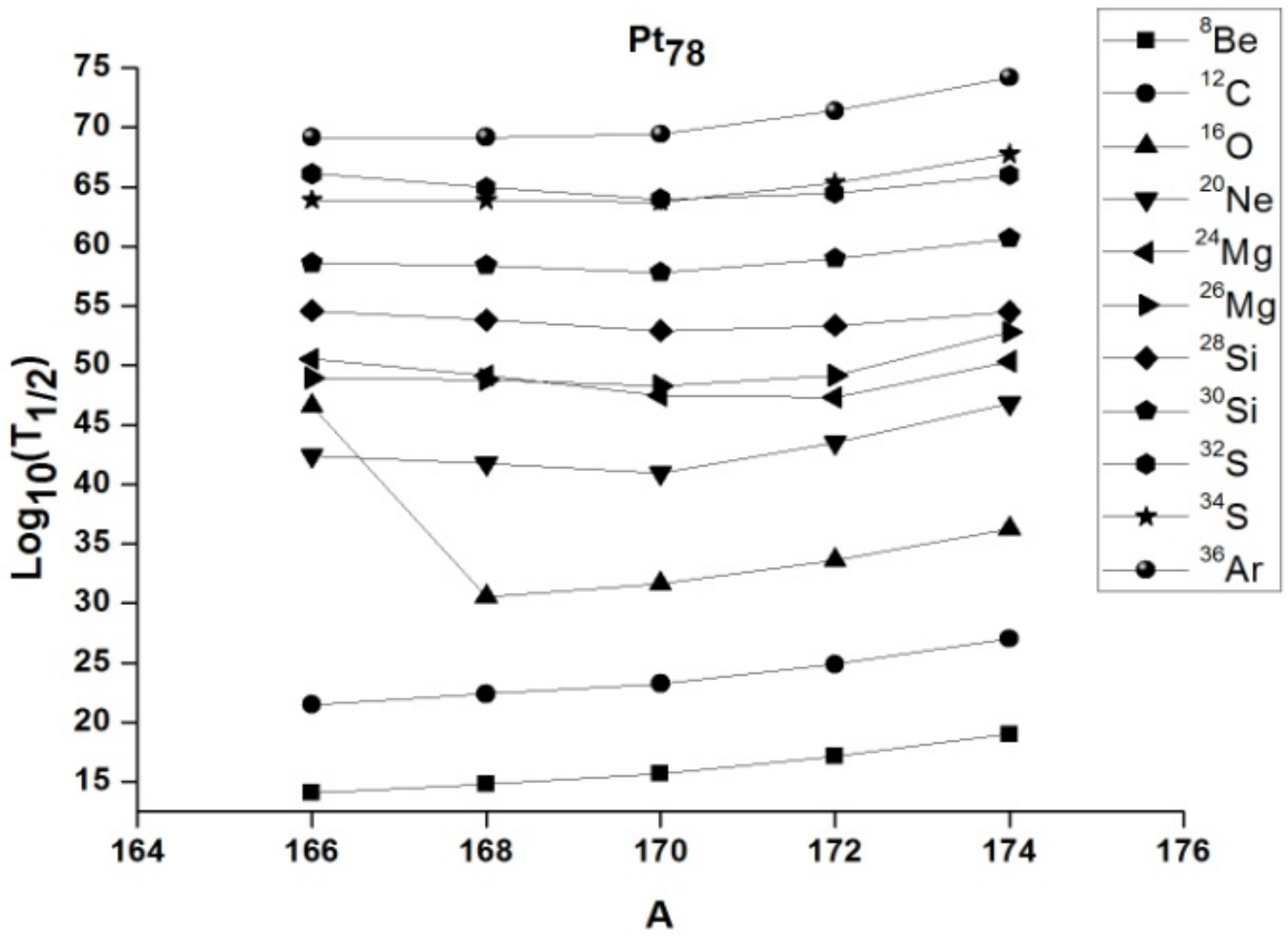

**Figure (2d) :- The values of $\log_{10}T_{1/2}$ versus the mass number (A) of parent isotopes $^{166\text{-}174}$Pt for different clusters.**

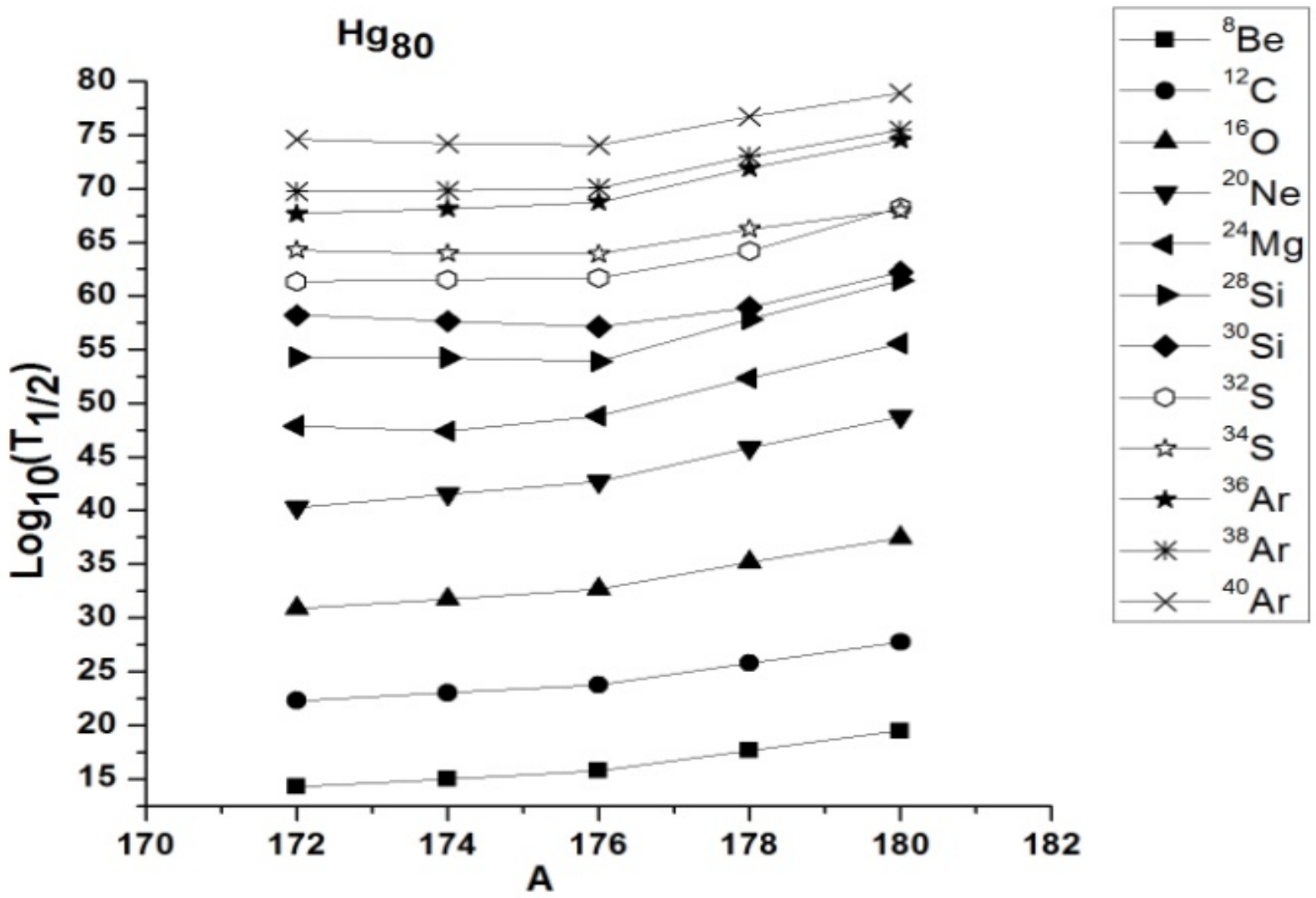

**Figure (2e) :- The values of $\log_{10}T_{1/2}$ versus the mass number (A) of parent isotopes $^{172\text{-}180}$Hg for different clusters**

The calculated values of $\log_{10}(T_{1/2})$ for alpha decay using CPPM for the chosen isotopes are plotted versus the mass number of the parent as shown in figure (3). This figure shows the calculated values compared with the experimental data [16]. It is obvious that, the isotopes of heavier atomic mass (A= 168 – 180) are more prone to alpha decay than the lighter ones (A= 156 – 166). Moreover, the daughters nuclei for the alpha-decay of the $^{156-162}$Hf are Yb isotopes with atomic number Z = 70 and neutron number N = 82, 84, 86 and 88 which implies the magic or near magic number of these neutron shells. The value of $\log_{10}(T_{1/2})$, is low for the $^{156}$Hf isotope due to N =82 closure in $^{152}$Yb daughter. Also, the daughter nuclei for alpha –decay of the $^{158-166}$W parent are the Hf isotopes with Z=72 and N= 84, 86, 88, 90 and 92 which near the magic nature of these neutron shells. The $^{162-170}$Os parents decay by alpha with W daughters with Z = 74 and N= 88 – 94. It is clear that alpha-decay half- lives have the lowest values for $^{166-174}$Pt and $^{172-180}$Hg isotopes. This can be explained as, the values of $\log_{10}(T_{1/2})$ decreasing due to the atomic and mass number increasing. Finally, figure (3) shows the calculated values of $\log_{10}(T_{1/2})$ are in less agreement with the experimental results [16] especially with increasing the atomic mass number of the parent nuclei.

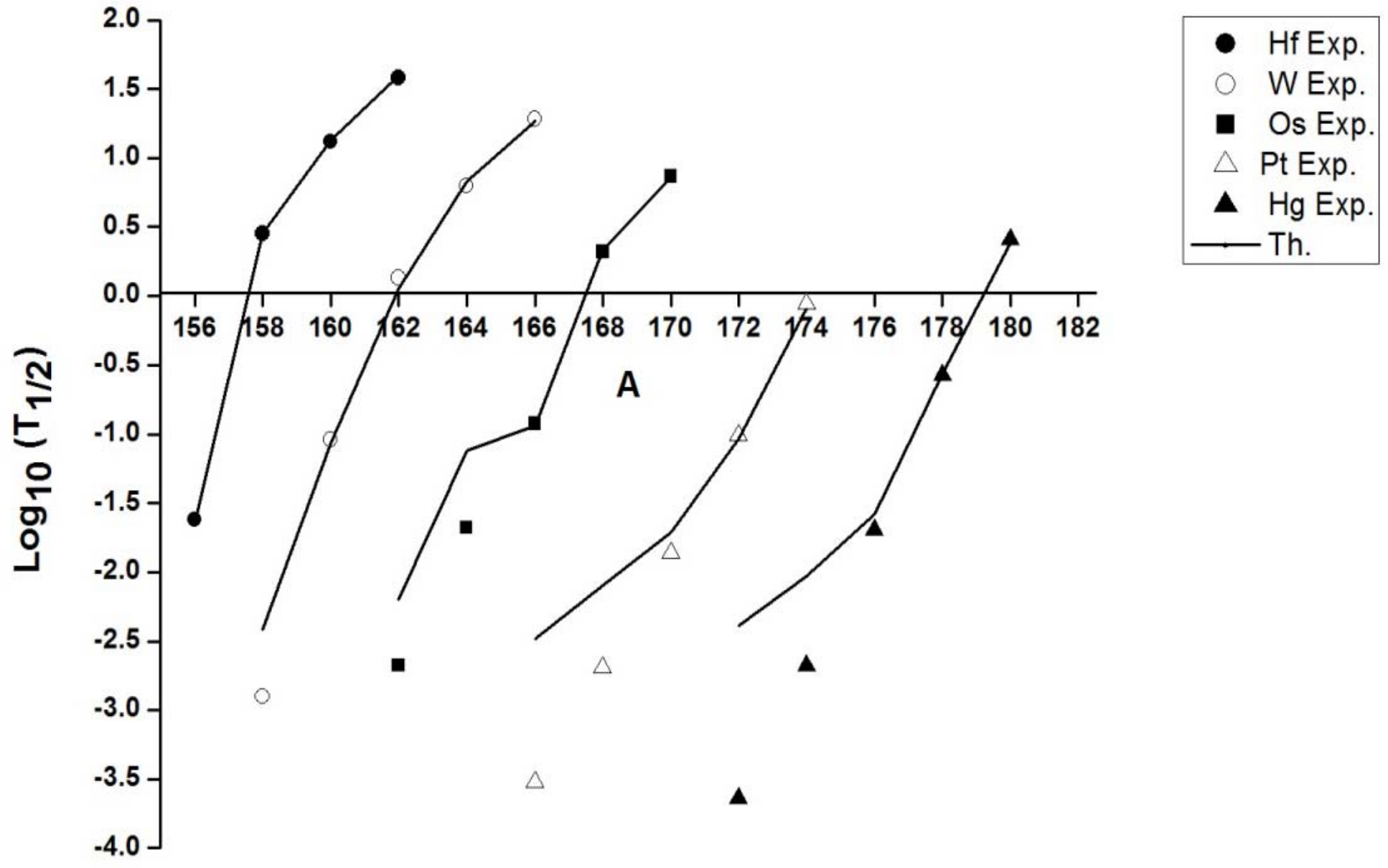

**Figure (3) :- The values of $\log_{10}T_{1/2}$ versus the mass number (A) of isotopes of the parent nuclei ($^{156-162}$Hf, $^{158-166}$W, $^{162-170}$Os, $^{166-174}$Pt, and $^{172-180}$Hg) compared with the corresponding experimental values for $\alpha$ -decay [16].**

## Conclusion

The cold valley (V-*Q*) plots for the $^{156-162}$Hf, $^{158-166}$W, $^{162-170}$Os, $^{166-174}$Pt and $^{172-180}$Hg isotopes are analyzed to determine the possible clusters emitted from these isotopes. From the cold valley plots, we noticed that the selection of the most probable clusters are depending on the shell closure for one of the emitted fragments or both of them. The half- lives of clusters decay are calculated using CPPM and UNIV and listed in table (1). The calculated values of $\log_{10}(T_{1/2})$ by CPPM are in agreement with the experimental values of alpha–decay more than the calculated values by the UNIV formula. The *Q*-value of the clusters have N = Z greater than of the clusters have N ≠ Z. From this study we conclude that, the isotopes with the greater mass number are exposed to disintegrate through alpha than the isotopes with smaller mass number. Hence, the half-life time of alpha decay decreases as the mass number of the parent increases.

## References

[1] A. Sandulescu, D. N. Poenaru and W. Greiner, Sov. J. Part. Nucl.**II**, 528 (1980).
[2] L Zheng, G. L. Zhang, J.C.Yang and W.W.Qu, Nucl. Phys. **A915**, 70 (2013).
[3] D. N. Basu, Phys. Lett . **B566**, 90 (2003).
[4] Z. Z. Ren, C. Xu, Z. J.Wang, Phys. Rev. **C70**, 034304 (2004).
[5] G. Royer, R. Moustabchir, Nucl. Phys.**A683**, 182 (2001).
[6] J. Blocki, J. Randrup, W. J. Swiatecki, C. F. Tsang, Ann. Phys.**(NY)105**, 427 (1977).
[7] W. D. Myers, W. J. Swiatecki, Phys. Rev. **C62**, 044610 (2000).
[8] P. Möller, J. R. Nix, Nucl. Phys. **A361**, 117 (1981).
[9] K. P. Santhosh, Antony Joseph, Pramana J. Phys. **58**, 611 (2002).
[10] K. P. Santhosh, Sabina Sahadevan, Nucl. Phys. **A847**, 42 (2010).
[11] R. K. Biju, Sabina Sahadevan, K. P. Santhosh, Antony Joseph, Pramana J. Phys.**70**, 427 (2008).
[12] K. E. Abd El Mageed and A. G. Shalaby, Chin. J. Phys. **53**, 040301 (2015), nucl-th/1312.2471.
[13] Y.J. Shi, W.J. Swiatecki, Nucl. Phys. **A438**, 450 (1985).
[14] K. P. Santhosh, Sabina Sahadevan, B. Priyanka, M. S. Unnikrishnan, Nucl. Phys. **A882**, 49 (2012).
[15] J. Blocki, W. J. Swiatecki, Ann. Phys. **(NY)132**, 53 (1981) .
[16] National Nuclear Data Center (NNDC) in Brookhaven National Laboratory, http://www.nndc.bnl.gov/
[17] D. N. Poenaru, M. Ivascu, A. Sandulescu, W. Greiner, Phys. Rev. **C32**, 572 (1985).
[18] D. N. Poenaru, R. A. Gherghescu, W. Greiner, Phys. Rev. **C83**, 014601 (2011).